\documentclass[12pt]{article}
\usepackage[height=8.5in,width=6.5in]{geometry}
\usepackage{hyperref}
\usepackage{tikz}
\usepackage{amssymb,amsmath}
\usepackage{comment}

\renewcommand{\tilde}{\widetilde}
\renewcommand{\bar}{\overline}

\newcommand{\cO}{\mathcal{O}}

\newcommand{\cJ}{\mathcal{J}}

\begin{document}
\setcounter{tocdepth}{2}

\begin{titlepage}
\begin{flushright}
YITP-16-129\\
\end{flushright}

\vskip2cm

\begin{center}
%{\A LARGE Note on Sachdev-Ye-Kitaev Like Model 
%without \\[4mm] Quenched Disorder}
{\LARGE A Note on Sachdev-Ye-Kitaev Like Model without
 \\[3mm] Random Coupling}

\vskip1.5cm
Takahiro Nishinaka\footnote{takahiro.nishinaka [at] yukawa.kyoto-u.ac.jp} and Seiji Terashima\footnote{terasima [at] yukawa.kyoto-u.ac.jp}
\vskip.5cm

{\it Yukawa Institute for Theoretical Physics,}\\[1mm]

{\it Kyoto University, Kyoto 606-8502, Japan}\\

\end{center}
\vskip2cm

\noindent
We study a description of the large $N$ limit of the Sachdev-Ye-Kitaev (SYK) model in terms of quantum mechanics without quenched disorder. Instead of random couplings, we introduce massive scalar fields coupled to fermions, and study a small mass limit of the theory. We show that, under a certain condition,
%the two, four and six point functions 
the correlation functions
of fermions reproduce those of the SYK model 
with a temperature dependent coupling constant
in the large $N$ limit. We also discuss a supersymmetric generalization of our quantum mechanical model. As a byproduct, we develop an efficient way of estimating the large $N$ behavior of correlators in the SYK model.

\end{titlepage}
\tableofcontents

\section{Introduction and summary}

%A quenched disorder is a random coupling constant in quantum field
%theory, and has widely been studied in condenced matter physics.
%%search for quantum mechanical models with large ground state degeneracy. In particular,
%Recently, 
The Sachdev-Ye-Kitaev (SYK) model \cite{Kitaev:2015, Sachdev:1992fk} is a quantum mechanical model with random coupling constants, and has recently attracted much attention for its possible connection to black holes \cite{Almheiri:2014cka, Sachdev:2015efa, Hosur:2015ylk, Polchinski:2016xgd, Michel:2016kwn, You:2016, Fu:2016yrv, Jevicki:2016bwu, Almheiri:2016fws, Jevicki:2016ito, Maldacena:2016hyu, Maldacena:2016upp, Danshita:2016xbo, Garcia-Alvarez:2016wem, Engelsoy:2016xyb, Jensen:2016pah, Bagrets:2016cdf, Cvetic:2016eiv, Gu:2016oyy, Berkooz:2016cvq, Garcia-Garcia:2016mno, Fu:2016vas, You:2016ldz, Witten:2016iux, Gurau:2016lzk, Cotler:2016fpe, Klebanov:2016xxf}. In particular, the model saturates the chaos bound proposed in \cite{Maldacena:2015waa} and therefore is expected to be dual to a black hole.

The random couplings are called ``quenched disorders,'' and follow the Gaussian distribution. Then the disordered correlation function of operators $\cO_1,\cdots,\cO_n$ is defined by
\begin{align}
\overline{\langle \cO_1\cdots \cO_n\rangle} \equiv \int dJ\, e^{-\alpha J^2} \frac{\int [d\psi]\cO_1\cdots \cO_n e^{-S[\psi,J]}}{\int [d\psi] e^{-S[\psi,J]}}~,
\label{eq:cor1}
\end{align}
where $J$ is the disorder and $\psi$ stands for quantum fields. The constant $\alpha$ is a parameter characterizing the Gaussian distribution of the disorder.\footnote{Here the integration measure $dJ$ is normalized so that $\langle 1 \rangle = 1$.} In the above definition, we first compute the correlation function for fixed $J$ and then take its average under the Gaussian distribution of $J$. Therefore it is not clear whether \eqref{eq:cor1} plays the same role as usual correlators of quantum field theories in various contexts such as the gauge/gravity correspondence.
In particular, if the SYK model describes %the
a black hole,
we need to understand the physical origin of the disorder.
Therefore, it may be desired to promote the SYK model
to %the 
an ordinary quantum mechanical model without disorders.

In this paper, we propose a quantum mechanical model without quenched disorders.
In the model, we just promote the variables $J$
into 
%the 
dynamical fields, 
which are %the 
free bosonic fields, i.e. harmonic oscillator, with small mass.
We show that our model reproduces essentially 
the large $N$ limit of the SYK model,
however, $\cJ$ should be replaced by a temperature dependent 
coupling $\cJ_{eff}$.\footnote{
Recently, in the paper \cite{Witten:2016iux},
another interesting model which reproduces
the large $N$ limit of the SYK model was given.} 
% in a limit of parameters.
Indeed, 
the large $N$ behavior of 
the two and four point correlators of the 
SYK model are reproduced 
by our model.
%thus 
This means that our model also 
%satisfies the saturation of 
saturates the chaos bound 
proposed in \cite{Maldacena:2015waa},
and is consistent with the effective action with the Schwarzian derivative discussed in
\cite{Maldacena:2016hyu}.
We note that the thermodynamic quantities including the free energy 
of our model are 
completely different from ones in the SYK model 
because of the almost massless bosons.
The correlators with %higher
larger numbers of external fields
are also reproduced in a sense explained below.
The simple replacement by the harmonic oscillators
behaves as the Gaussian quenched disorder
due to some special properties 
of the large $N$ limit of the SYK model.

An important previous work in this direction
is \cite{Michel:2016kwn}
where %the H
harmonic oscillators %was
are introduced
and the random couplings $\cJ$ are replaced by the momenta of the %H
harmonic oscillators.
We discuss the path-integral formulation of this model 
is essentially described by the action of our model.

We also propose a supersymmetric generalization of our model
which reproduces the correlators of the supersymmetric generalization of the SYK model \cite{Fu:2016vas}.
Here the super-partners of the almost massless bosons are decoupled 
from others in the large $N$ limit.

One of the motivations for the study of the SYK model is its relevance to the black hole physics. If our model indeed has some application to the black hole physics,
the %${\cal O}(N^4)$ 
almost massless bosons 
should be interpreted appropriately.
Although it is unclear that our model describes
black hole physics or not,
we hope that our field theory analogue of the 
SYK model will be important for further study 
of the black hole physics and other areas.

The organization of this paper is the following. In section 2, we review the SYK model and also provide an efficient way of estimating the large $N$ behavior of correlators. In section 3, we propose a quantum mechanical model without quenched disorders and argue that our model reproduces important properties of the SYK model in the large $N$ limit. In section 4, we discuss a supersymmetric generalization.

\section{SYK model}

The SYK model is a quantum mechanics of $N$ Majorana fermions, $\psi_i$
for $i=1,\cdots,N$, with quenched disorders. The Hamiltonian of the
model\footnote{
This can be considered as a special case with $q=4$ of the 
generalized SYK model \cite{Maldacena:2016hyu}
where the number of the fermions in the Hamiltonian is $q$.
Our considerations in this paper can be easily generalized
to other cases of $q$ %case 
although we do not do it explicitly. 
}
is given by
\begin{align}
H = \sum_{1\leq i<j<k<\ell\leq N} J_{ijk\ell}\, \psi_{i}\psi_{j}\psi_{k}\psi_{\ell}~,
\label{eq:Hamiltonian}
\end{align}
where $J_{ijk\ell}$ are disorders following the Gaussian distribution such that
\begin{align}
\bar{J_{ijk\ell}} = 0~,\qquad \bar{(J_{ijk\ell})^2} = \frac{3!}{N^3}\cJ^2~.
\end{align}
The constant $\cJ$ has dimension of mass, and therefore becomes large in the infrared.
Below we will work in Euclidean space.
The Lagrangian corresponding to the Hamiltonian \eqref{eq:Hamiltonian} is written as
\begin{align}
L(\psi;J) = -\frac{1}{2}\sum_{i=1}^N \psi_i \frac{d}{d\tau} \psi_i - H~,
\label{eq:Lagrangian}
\end{align}
and the disordered correlation function of $n$ fermions is given by
\begin{align}
\overline{\langle \psi_{i_1}(\tau_1)\cdots\psi_{i_n}(\tau_n)\rangle} = \int \left(\prod_{i,j,k,l}d J_{ijk\ell}\; e^{-\alpha (J_{ijk\ell})^2}\right) \frac{\int [d\psi] \psi_{i_1}(\tau_1)\cdots\psi_{i_n}(\tau_n) e^{-\int d\tau L}}{\int [d\psi] e^{-\int d\tau L }}~,
\label{eq:correlator2}
\end{align}
where $\alpha \equiv N^3/12\cJ^2$. 
Explicit computations of two and four point functions of this model were discussed in \cite{Kitaev:2015, Polchinski:2016xgd, Maldacena:2016hyu}.

%Note here that the correlation function \eqref{eq:correlator2} is different from
%\begin{align}
%\frac{\int dJ_{ijk\ell} [d\psi]\, \psi_{i_1}(t_1)\cdots \psi_{i_n}(t_n)\; e^{-\alpha (J_{ijk\ell})^2-\int dt\, L}}{\int dJ_{ijk\ell} [d\psi] e^{-\alpha (J_{ijk\ell})^2 - \int dt\, L}}~.
%\label{eq:correlator3}
%\end{align}
%where $J_{ijk\ell}$ is regarded as a constant auxiliary field instead of
%a random coupling. Note that this constant field induces non-local fields and does not give usual one-dimensional field theory.

%%However, we will show later that
%%\eqref{eq:correlator3} coincides with \eqref{eq:correlator2} in the
%%large $N$ limit with $\cJ$ fixed. \note{Takahiro}{This paragraph might
%%have to be removed.}

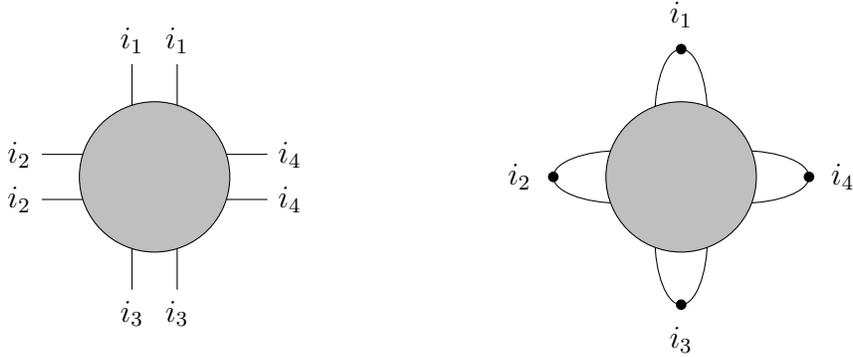
\begin{figure}
\begin{center}
\vskip .6cm
\begin{tikzpicture}
\draw (-0.3,-1.5) node[below]{$i_3$} -- (-0.3,1.5) node[above]{$i_1$};
\draw (0.3,-1.5) node[below]{$i_3$} -- (0.3,1.5) node[above]{$i_1$};
\draw (-1.5,0.3) node[left]{$i_2$} -- (1.5,0.3) node[right]{$i_4$};
\draw (-1.5,-0.3) node[left]{$i_2$} -- (1.5,-0.3) node[right]{$i_4$};
\draw[fill=gray!50] (0,0) circle [radius=1];

\draw (7,0.8) ellipse (0.35cm and 0.9cm);
\fill (7,1.7) node[above=1.5mm]{$i_1$} circle (2pt);
\draw (7.8,0) ellipse (0.9cm and 0.35cm);
\fill (8.7,0) node[right=1.5mm]{$i_4$} circle (2pt);
\draw (7,-0.8) ellipse (0.35cm and 0.9cm);
\fill (7,-1.7) node[below=1.5mm]{$i_3$} circle (2pt);
\draw (6.2,0) ellipse (0.9cm and 0.35cm);
\fill (5.3,0) node[left=1.5mm]{$i_2$} circle (2pt);
\draw[fill=gray!50] (7,0) circle [radius=1];
\end{tikzpicture}
\caption{Two expressions for external lines in the diagram. The shaded region is arbitrary. \; Left: The conventional one.\; Right: Our new expression
for the same diagram in terms of dots. Each dot expresses two external fermions in a pair in the sense explained in the main text. Since we take the sum over $i_k$, we usually omit the indices in the diagram.}
\label{fig:circle}
\end{center}
\end{figure}

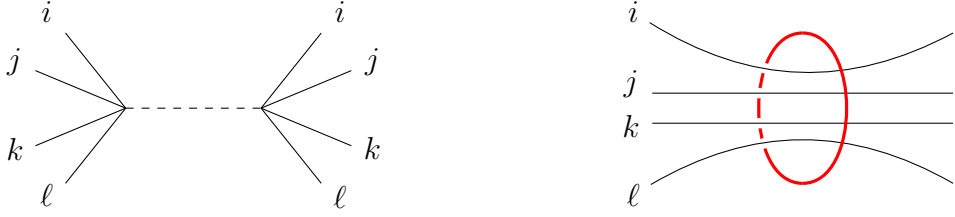
\begin{figure}
\begin{center}
\vskip .6cm
\begin{tikzpicture}
\draw (0,0) -- (-.8,1) node[left=2.5mm,above]{$i$};
\draw (0,0) -- (-1.2,.5) node[left=2.8mm,above=-2mm]{$j$};
\draw (0,0) -- (-1.2,-.5) node[left=2.7mm,below=-2.5mm]{$k$};
\draw (0,0) -- (-.8,-1) node[left=2.5mm,below=-1.3mm]{$\ell$};
\draw[dashed] (0,0) -- (1.8,0);
\draw (1.8,0) -- (2.6,1) node[right=2.5mm,above]{$i$};
\draw (1.8,0) -- (3,.5) node[right=2.8mm,above=-2mm]{$j$};
\draw (1.8,0) -- (3,-.5) node[right=2.8mm,below=-2.6mm]{$k$};
\draw (1.8,0) -- (2.6,-1) node[right=2.5mm,below=-1.1mm]{$\ell$};
\path (9,1) edge[bend right=90, color=red, very thick] (9,-1);
\fill[color=white] (8.5,.53) circle (2pt);
\fill[color=white] (8.43,0.2) circle (2pt);
\fill[color=white] (8.43,-0.2) circle (2pt);
\fill[color=white] (8.48,-.47) circle (2pt);
\path (7,1) node[left=2.5mm,above]{$i$} edge[bend right] (11,1);
\draw (7,0.2) node[left=2.8mm,above=-2mm]{$j$} -- (11,0.2);
\draw (7,-0.2) node[left=2.8mm,below=-2.5mm]{$k$} -- (11,-0.2);
\path (7,-1) node[left=2.5mm,below=-1.3mm]{$\ell$} edge[bend left] (11,-1);
\path (9,1) edge[bend left=90, color=red, very thick] (9,-1);
\end{tikzpicture}
\caption{Two expressions for an interaction vertex. Left: The conventional one. The dashed line stands for the disorder average associated with $J_{ijk\ell}$. \; Right: Another expression for the left picture in terms of a ``bundle.'' The red circle in the middle stands for a bundle corresponding to the disorder average in the left picture. We say two fermion lines are connected to each other when they are involved in the same bundle. }
\label{fig:vertex}
\end{center}
\end{figure}

Let us illustrate how to read off the large $N$ behavior of correlation functions in the SYK model. As in \cite{Polchinski:2016xgd, Maldacena:2016hyu}, 
%let us take the large $N$ limit and consider the 
we focus on correlators of the following form%only
:
\footnote{
We will not consider other types of correlators in this paper
although they might be important.}
\begin{eqnarray}
\sum_{i_1,i_2, \cdots, i_{n}=1}^N 
\overline{\langle 
\psi_{i_1}(\tau_1) \psi_{i_1}(\tau_2) 
\cdots \psi_{i_n}(\tau_{2n-1} ) \psi_{i_n}(\tau_{2n})
\rangle}~.
\label{cf1}
\end{eqnarray}
%Bcause of this pair-wise form,
Since the subscripts of the fermions are in pairs, it is useful to divide external lines in Feynman diagrams into the corresponding pairs. We then connect external lines in each such pair with a ``dot'' (fig.~\ref{fig:circle}). While this operation obscures the locations of the external fermions in time, it makes it easy to read off the large $N$ behavior of the contribution from each diagram. In the rest of this paper, a ``dot'' always stands for two external fermions in a pair in this sense.
%we will denote the each pair of the external lines connected each other with an insersion of a dot in the propagator line. (**** put fig 1).
%We regard this dotted line as a connected line
%although this represents two external lines.
After
%integrating out 
the integration over the disorders $J_{ijk\ell}$,
the fermion fields $\psi_i$ have only the eight-point interaction of the form
%there is only the eight-points vertex which 
%is the following form:
\begin{align}
\frac{3!\cJ^2}{N^3}\int d\tau_1d\tau_2
\sum_{1\leq i_1<i_2<i_3<i_4<N} \psi_{i_1}(\tau_1) 
\psi_{i_1}(\tau_2) 
\psi_{i_2}(\tau_1) 
\psi_{i_2}(\tau_2) 
\psi_{i_3}(\tau_1) 
\psi_{i_3}(\tau_2) 
\psi_{i_4}(\tau_1) 
\psi_{i_4}(\tau_2)~.
\end{align}
%This interaction vertex is denoted
We denote this interaction vertex by a ``bundle'' of four lines
%%where
to which the indices $i_1,i_2,i_3,i_4$ are assigned
% to the four lines
(fig.~\ref{fig:vertex}).
%In the rest of this paper, we call a set of bundled four lines a ``bundle.''
This expression again obscures the locations of fermions in time, but enables us to read off the large $N$ behavior easily.

%Thus,
In terms of the above ``dot'' and ``bundle,'' any Feynman diagram %is
can be expressed as a collection of circles with dots and bundles attached to them, where each circle carries $N$ degrees of freedom.
%bundled by the vertex because 
Note that there are neither branch points
%and 
nor end points of lines %in the lines.
in this expression. In the rest of this paper, we use this new expression for diagrams.
We will denote the number of %the
 circles by $L$, and that of bundles by $V$.
%, which need not coincide with the number of the loops in the usual sense,
 Note that $L$ needs not coincide with the number of fermion loops in the usual expression for the diagram, while $V$ equals to the number of vertices in the usual expression.
%With this, the $N$ dependence of the diagram is 
Then, the contribution from the diagram is of
${\cal O} (N^{L-3V})$.
Note that this does not depend on the number of the dots in the diagram
%which represent the external lines. 
(i.e. (half) the number of external lines in the usual expression for the diagram).

Now, we can show that
for connected diagrams
the leading order %behavior
contribution is always of ${\cal O} (N)$
irrespective to the number of %the
dots in the diagram.
Indeed, for $V=0$, the only connected diagram is 
a circle with dots, which has a summation over only an index
%, thus gives a factor $N$.
and therefore is of $\cO(N)$.
%With
On the other hand, from $k$ diagrams of $\cO(N)$, we can generate 
a new connected diagram by bundling circles.\footnote{ %by a vertex. 
This operation corresponds to inserting an additional vertex in the usual expression for the diagram.}
The bundle is associated with an extra factor of $\cJ^2/N^3$, and therefore the new diagram is of ${\cal O}(N^{k-3})$. Since $1\leq k \leq 4$, the leading contribution is again of $\cO(N)$.
%the diagrams which are bundled by the vertex.
%Of course, $n$ should satisfy $1\le n\leq 4$,
%which implies the leading diagrams are of ${\cal O} (N)$ by the mathematical induction.
Since any connected diagram can be generated by repeating this procedure, we see that the leading contribution is always of $\cO(N)$ for connected diagrams.
This also implies that
the leading order (connected) diagram has the following property:
if we %remove
``untie'' any 
%vertex (bundle) 
bundle 
%in
from the diagram, 
%then the resulting diagrams have
the diagram splits into four disconnected parts (fig.~\ref{fig:disconnected}).
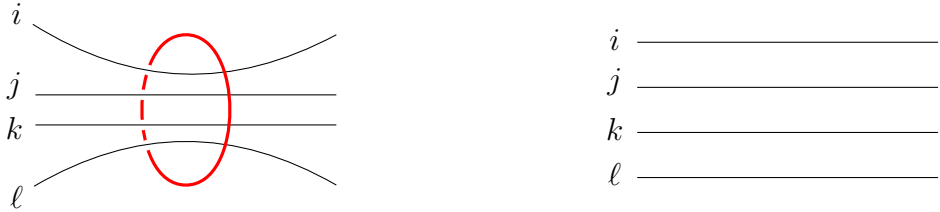
\begin{figure}
\begin{center}
\vskip .6cm
\begin{tikzpicture}
\path (1,1) edge[bend right=90, color=red, very thick] (1,-1);
\fill[color=white] (0.5,.53) circle (2pt);
\fill[color=white] (0.43,0.2) circle (2pt);
\fill[color=white] (0.43,-0.2) circle (2pt);
\fill[color=white] (0.48,-.47) circle (2pt);
\path (-1,1) node[left=2.5mm,above]{$i$} edge[bend right] (3,1);
\draw (-1,0.2) node[left=2.8mm,above=-2mm]{$j$} -- (3,0.2);
\draw (-1,-0.2) node[left=2.8mm,below=-2.5mm]{$k$} -- (3,-0.2);
\path (-1,-1) node[left=2.5mm,below=-1.3mm]{$\ell$} edge[bend left] (3,-1);
\path (1,1) edge[bend left=90, color=red, very thick] (1,-1);

\draw (7,0.9) node[left=3mm,above=-2.5mm]{$i$} -- (11,0.9);
\draw (7,0.3) node[left=3mm,above=-2.5mm]{$j$} -- (11,0.3);
\draw (7,-0.3) node[left=3mm,above=-2.5mm]{$k$} -- (11,-0.3);
\draw (7,-0.9) node[left=3mm,above=-2.5mm]{$\ell$} -- (11,-0.9);
\end{tikzpicture}
\caption{``Untying'' a bundle corresponds to replacing the left picture with right. }
\label{fig:untying}
\end{center}
\end{figure}
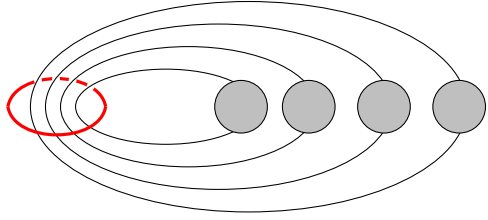
\begin{figure}[h]
\begin{center}
\vskip .6cm
\begin{tikzpicture}
\path (-1.1,0) edge[bend left=80, color=red, very thick] (.2,0);
\fill[color=white] (-0.7,0.35) circle (1.5pt);
\fill[color=white] (-0.46,0.37) circle (1.5pt);
\fill[color=white] (-0.24,0.35) circle (1.5pt);
\fill[color=white] (0,0.27) circle (1.5pt);
\draw (1,0) ellipse (1.2cm and 0.5cm);
\draw (1.3,0) ellipse (1.7cm and 0.8cm);
\draw (1.7,0) ellipse (2.3cm and 1.1cm);
\draw (2.1,0) ellipse (2.9cm and 1.4cm);
\draw[fill=gray!50] (2,0) circle [radius=.35];
\draw[fill=gray!50] (2.9,0) circle [radius=.35];
\draw[fill=gray!50] (3.9,0) circle [radius=.35];
\draw[fill=gray!50] (4.9,0) circle [radius=.35];
\path (-1.1,0) edge[bend right=80, color=red, very thick] (.2,0);
\end{tikzpicture}
\caption{A diagram whose contribution is of $\cO(N)$ splits into four disconnected parts when a bundle in it is untied. Here the shaded regions are arbitrary.}
\label{fig:disconnected}
\end{center}
\end{figure}
\begin{figure}[h]
\begin{center}
\vskip .6cm
\begin{tikzpicture}
\path (-1.1,0) edge[bend left=80, color=red, very thick] (.2,0);
\fill[color=white] (-0.7,0.35) circle (1.5pt);
\fill[color=white] (-0.46,0.37) circle (1.5pt);
\fill[color=white] (-0.24,0.35) circle (1.5pt);
\fill[color=white] (0,0.27) circle (1.5pt);
\draw (1,0) ellipse (1.2cm and 0.5cm);
\draw (1,0) ellipse (1.4cm and 0.7cm);
\draw (1,0) ellipse (1.6cm and 0.9cm);
\draw (1,0) ellipse (1.8cm and 1.1cm);
\draw[fill=gray!50] (2.5,0) circle [radius=1.1];
\path (-1.1,0) edge[bend right=80, color=red, very thick] (.2,0);
\draw (9,0) -- (6.9,1.1);
\draw (9.4,0) -- (6.95,1.25);
\draw (8.7,0) -- (6.85,0.95);
\draw (8.4,0) -- (6.8,0.8);
\draw (9,0) -- (6.9,-1.1);
\draw (9.4,0) -- (6.95,-1.25);
\draw (8.7,0) -- (6.85,-0.95);
\draw (8.4,0) -- (6.8,-0.8);
\draw[fill=gray!50] (8.5,0) circle [radius=1.1];
\end{tikzpicture}
\caption{ ``Cutting'' a bundle in the diagram implies replacing the left picture with the right. This corresponds to cutting the dashed line in fig.~\ref{fig:vertex}. Note that this ``cutting'' is different from ``untying'' a bundle discussed in fig.~\ref{fig:untying}.}
\label{fig:cutting}
\end{center}
\end{figure}
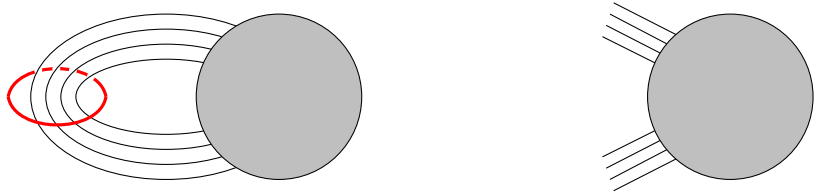
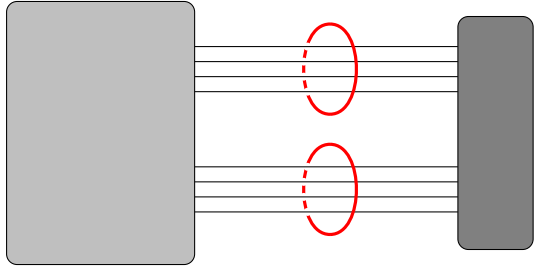
\begin{figure}[h]
\begin{center}
\vskip .6cm
\begin{tikzpicture}
\path (3.3,3.2) edge[bend right=90, color=red, very thick] (3.3,2);
\fill[color=white] (3,2.9) circle (1.5pt);
\fill[color=white] (2.95,2.7) circle (1.5pt);
\fill[color=white] (2.95,2.5) circle (1.5pt);
\fill[color=white] (3,2.3) circle (1.5pt);

\path (3.3,1.6) edge[bend right=90, color=red, very thick] (3.3,0.4);
\fill[color=white] (3,1.3) circle (1.5pt);
\fill[color=white] (2.95,1.1) circle (1.5pt);
\fill[color=white] (2.95,0.9) circle (1.5pt);
\fill[color=white] (3,0.7) circle (1.5pt);

\draw (1.5,2.9) -- (5,2.9);
\draw (1.5,2.7) -- (5,2.7);
\draw (1.5,2.5) -- (5,2.5);
\draw (1.5,2.3) -- (5,2.3);

\draw (1.5,1.3) -- (5,1.3);
\draw (1.5,1.1) -- (5,1.1);
\draw (1.5,0.9) -- (5,0.9);
\draw (1.5,0.7) -- (5,0.7);

\draw[fill=gray!50, rounded corners] (-1,0) rectangle (1.5,3.5);
\draw[fill=gray, rounded corners] (5,0.2) rectangle (6,3.3);

\path (3.3,3.2) edge[bend left=90, color=red, very thick] (3.3,2);
\path (3.3,1.6) edge[bend left=90, color=red, very thick] (3.3,0.4);

\end{tikzpicture}
\caption{A diagram which produces a ``bubble'' when enough number of bundles are cut. The right shaded region contains no ``dot'' while the left shaded region is arbitrary.}
\label{fig:bubble}
\end{center}
\end{figure}

%Now, let us consider 
We conclude this section by comparing the disordered correlator \eqref{eq:correlator2} with the following quantity:\footnote{
This consideration was essentially given in \cite{Michel:2016kwn}.} 
\begin{align}
\frac{\int dJ_{ijk\ell} [d\psi]\, \psi_{i_1}(\tau_1)\cdots \psi_{i_n}(\tau_n)\; e^{-\alpha (J_{ijk\ell})^2-\int d\tau\, L}}{\int dJ_{ijk\ell} [d\psi] e^{-\alpha (J_{ijk\ell})^2 - \int d\tau\, L}}~,
\label{eq:correlator3}
\end{align}
where $J_{ijk\ell}$ is regarded as a constant auxiliary field instead of a random coupling. 
%the difference between the correlator with the disorder\eqref{eq:cor1}
%and the one in \eqref{eq:correlator3}.
%Only the difference is in the denominators
%which remove the vacuum diagrams.
The only difference between \eqref{eq:correlator2} and \eqref{eq:correlator3} is the notion of vacuum bubbles. To see this, let us consider connected diagrams with the following property:
%(*** fig~\ref{fig:property}):
%We will consider 
%connected diagrams with the following property (**** fig 5):
%if we cut the diagram at some of vertices,
if we ``cut'' enough number of bundles in the diagram, as in fig.~\ref{fig:cutting}, then the resulting diagram has at least one connected component without
%the 
dots in it (fig.~\ref{fig:bubble}).\footnote{Recall here that a ``dot'' stands for two external fermions in a pair in the sense explained in fig.~\ref{fig:circle}, and also that ``cutting'' a bundle means cutting the dashed line (i.e., a propagator of $\phi$) in the sense described in fig.~\ref{fig:vertex}.}
% which is disconnected from others.
%This component in the diagram is regarded as a vacuum bubble
We call such a connected component without dots a ``bubble'' because it is regarded as a vacuum bubble in the computation of \eqref{eq:correlator2}. Then we denote by $n_B$ the (minimum) number of bundles we should cut to separate as many bubbles as possible from the diagram. Any diagram producing a bubble in the above cutting operation does not contribute to the disordered correlator \eqref{eq:correlator2} while it contributes to the quantity \eqref{eq:correlator3}.\footnote{
Note that this bubble diagram corresponds to 
an effective vertex or an effective propagator of $\phi$.}
%in the computations of the correlator in the disorder
%(\ref{eq:correlator2})
%because we should regard $J$ as coupling constants
%and then inetgrating out in the computations.
%
%Let us consider how the $N$ dependence changes 
%by replacing a vertex next to the vacuum bubble
%by four lines without vertices 
%in the original diagram (****  fig A).
%We will concentrate on 
%four lines emnating from the removed vertex
%into the vacuum bubble.
%If the original connected diagram fig. 5 is the leading order, i.e. ${\cal O}(N)$,
%it should be a diagram represented in fig. 4. 
%Thus, any one of the four lines in the 
%new diagram fig. 6 should be decoupled from other three lines
%and is given by (**** fig. B).
%However, this diagram is not possible because all lines should be circles.
However, we will see below that 
%there are diagrams which dominate
%these diagrams with the bubble.******
this difference between \eqref{eq:correlator2} and \eqref{eq:correlator3} is sub-leading in the large $N$ limit.

To see this, let us consider a diagram containing bubbles.
%let us cut bundles so that as many bubbles as possible are separated from the diagram. We denote by $n$ the minimum number of bundles we should cut in this process. We assume that at least one bubble is separated from the diagram in the process.
Then $n_B$ is greater than one
%the bundles (i.e. vertices) which next to attached to a bubble.
%as $n$, which 
%The integer $n$ should be greater than one
because of the anti-commutativity of the fermions.
Note that %any 
every line in one of the $n_B$ bundles %should be 
is connected, through a bubble,
%in the bubble 
to a line in another one of the $n_B$ bundles.
%Thus
This means that, if we %remove
``untie'' all the $n_B$ bundles,
%(i.e. vertices) which next to the bubble, the resulting diagram has 
the diagram splits into at most $2n_B$ disconnected parts.
%This means that the $N$ dependence of 
Since each such disconnected part is at most of $\cO(N)$, the original diagram is at most of order $N^{2n_B-3n_B}=N^{-n_B} \leq N^{-2}$.
Therefore, in comparison to the $\cO(N)$ leading contribution, %there are 
contributions from diagrams containing a bubble are
%(\ref{eq:cor1}) and (\ref{eq:correlator3}) %, however, in the large $N$ limit the differences are 
 sub-leading at least by the factor $N^{-3}$. This implies that \eqref{eq:correlator2} and \eqref{eq:correlator3} are identical at the leading order of $N$.
%smaller than the leading order diagrams and then negligible.

Note, however, that the constant field $J_{ijk\ell}$ in \eqref{eq:correlator3} induces non-local
%fields
interactions and therefore \eqref{eq:correlator3} cannot be interpreted as a correlator of a one-dimensional local quantum field theory. In the next section, we propose another quantum mechanical model to overcome this difficulty.

\section{Our model}

Let us consider the quantum mechanics of $\phi_{ijk\ell}(\tau)$ and
$\psi_i(\tau)$ described by the following action:
\begin{align}
\tilde{S}[\psi,\phi] = \int d\tau \sum_{i<j<k<\ell}
\frac{1}{2\epsilon}
\left(
\left(\frac{d\phi_{ijk\ell}}{d\tau} \right)^2 + m^2  (\phi_{ijk\ell})^2\right) + \int d\tau\, L(\psi,J=\phi(\tau))~,
\label{a1}
\end{align}
where $m$ is a mass parameter, $\epsilon/m^3$ is a dimensionless constant and $L(\psi,J)$ is the
Lagrangian defined in \eqref{eq:Lagrangian}
where the constant $J$ is replaced by the dynamical bosonic fields
(harmonic oscillators) $\phi(\tau)$. 
We define the correlation function in this theory as
\begin{align}
\langle \psi_{i_1}(\tau_1)\cdots \psi_{i_n}(\tau_n)\rangle \equiv 
\frac{\int [d\phi][d\psi] \;\psi_{i_1}(\tau_1)\cdots \psi_{i_n}(\tau_n)\; e^{-\tilde{S}[\psi,\phi]}}{\int [d\phi][d\psi] e^{-\tilde{S}[\psi,\phi]}}~,
\label{eq:correlator4}
\end{align}
and only consider the combinations of correlation functions of the form in (\ref{cf1}).
Below we will show that, for a range of $\epsilon$ and $m$, 
this correlation function almost agrees with \eqref{eq:correlator2} in the
large $N$ limit. 

Here we assume %that a
the small mass limit
\begin{eqnarray}
 m \ll |\omega_i|,
\end{eqnarray}
where $\omega_i$ is a momentum (or frequency)
of an external fermion line, and
the large $N$ limit is taken before 
%taking 
this limit.

\subsection{Correlation function}

\begin{figure}
\begin{center}
\vskip .6cm
\begin{tikzpicture}
%\path (-1.1,0) edge[bend left=80, color=red, very thick] (.2,0);
%\fill[color=white] (-0.7,0.35) circle (1.5pt);
%\fill[color=white] (-0.46,0.37) circle (1.5pt);
%\fill[color=white] (-0.24,0.35) circle (1.5pt);
%\fill[color=white] (0,0.27) circle (1.5pt);
%\draw (1,0) ellipse (1.2cm and 0.5cm);
%\draw (1,0) ellipse (1.4cm and 0.7cm);
%\draw (1,0) ellipse (1.6cm and 0.9cm);
%\draw (1,0) ellipse (1.8cm and 1.1cm);
%\path (-1.1,0) edge[bend right=80, color=red, very thick] (.2,0);
%\path (-1,-1.67) edge[bend left=50] (-0.3,0);
%\path (-1,-1.67) edge[bend left=75] (-0.7,0);
%\path (-1,-1.67) edge[bend left=80] (-1.25,-0.3);
%\draw (-1,-1.67) edge[bend left=90] (-1.5,0);
%%
%\path (1,-1.67) edge[bend right=50] (0.3,0);
%\path (1,-1.67) edge[bend right=75] (0.7,0);
%\path (1,-1.67) edge[bend right=80] (1.25,-0.3);
%\draw (1,-1.67) edge[bend right=90] (0,0);
\draw (1,-1.67) arc (20:160:1.07cm);
\draw (1,-1.67) arc (-21:201:1.08cm);
\draw (1,-1.67) arc (-43:223:1.37cm);
\draw (1,-1.67) arc (-53:233:1.68cm);
\draw[fill=gray!50] (0,1.4) circle [radius=.3];
\draw[fill=gray!50] (0,0.6) circle [radius=.3];
\draw[fill=gray!50] (0,-0.2) circle [radius=.3];
\draw[fill=gray!50] (0,-1) circle [radius=.3];
\draw[dashed] (1,-1.67)--(-1,-1.67) node[midway, below=1mm]{$\omega$};
%
%\draw[fill=gray!50] (0,0) ellipse (2.5cm and 1cm);
\end{tikzpicture}
\caption{Another expression for the diagram in in fig.~\ref{fig:disconnected}. Here we explicitly show a propagator of $\phi$ in terms of the dashed line. The shaded regions are arbitrary.}
\label{fig:leading}
\end{center}
\end{figure}
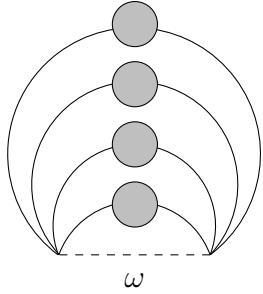
As we have explicitly seen in the previous section,
the relevant diagrams in the large $N$ limit in this model %which are relevant in the large $N$ limit
are the same as %the ones
those in the SYK model.\footnote{To be precise, the discussion in the previous section was on the large $N$ equivalence of \eqref{eq:correlator2} and \eqref{eq:correlator3}. While \eqref{eq:correlator3} and \eqref{eq:correlator4} are different, the same set of diagrams is relevant in the large $N$ limit. The difference appears in the contribution from each diagram, as discussed below.} In particular, the leading contributions from connected diagrams are of $\cO(N)$, and an interaction vertex in them can always be expressed as in fig.~\ref{fig:leading}.
%For any vertex in one of these diagrams
%it is represented as fig.4. or (**** fig.6) 
%The interaction vertex is the one in fig.~\ref{fig:vertex}, in the left of which the $\phi$ propagator is explicitly shown. This 
We here denote by $\omega$ the momentum (or frequency) of the $\phi$ propagator shown in fig.~\ref{fig:leading}, and consider the integral over $\omega$. Such an integral does not appear when every fermion line involved in the bundle is connected to a dot, because in that case $\omega$ is fixed by external momenta. We discuss such a case in the next subsection.
%the 
%any one of 
%each of the four fermion lines 
%contains %the 
%at least one dots, i.e., is connected to at least one external fermion.
%which is represented as an insersion of a dot on the line.
Clearly, such a diagram without the integration over $\omega$
contains $2k$ external lines %where 
for $k \ge 4$.
Thus, for two, four and six point correlators,
 $\omega$ is not fixed by external momenta. %are no such exceptional diagrams.

Let us first focus on two, four and six point correlators. %Then $\omega$ is not fixed by external momenta.
%For evaluating 
Then the contribution of the diagram %represented
shown in fig.~\ref{fig:leading} contains an integral over $\omega$.
We first split the integrand into the $\phi$ propagator, $\frac{\epsilon}{\omega^2+m^2}$, and the other part, $A(\omega,\omega_i)$, so that the contribution from the diagram is written as
\begin{align}
\epsilon \int_{-\infty}^{\infty}d\omega \frac{1}{\omega^2 + m^2} A(\omega,\omega_i)~.
\label{eq:integral}
\end{align}
Then we introduce an auxiliary intermediate scale $\Lambda$,
\begin{eqnarray}
 m \ll \Lambda \ll |\omega_i|,
\end{eqnarray}
and %we will
splits the integral \eqref{eq:integral} into two parts as
%\footnote{
%If the coupling is $\dot\phi \psi\psi\psi\psi$ 
%instead of $\phi \psi\psi\psi\psi$, the contribution
%is 
%\begin{align}
%  \epsilon \int_{-\infty}^{\infty}
%d \omega \frac{\omega^2}{\omega^2+m^2}  A(\omega,\omega_i)
%\approx   
%\epsilon \int_{-\infty}^{\infty} A(\omega,\omega_i),
%\end{align}
%which is clearly different from the SYK result
%where only $A(\omega=0,\omega_i)$ appears.
%} 
% separate the contributions of the diagram
%in the $\phi$ propagator and the other part which is denoted as
% $A(\omega,\omega_i)$.
%Then the contribution of the diagram is
\begin{align}
% \epsilon \int_{-\infty}^{\infty}
%d \omega \frac{1}{\omega^2+m^2}  A(\omega,\omega_i)
%=
 2 \epsilon \int_{0}^{\Lambda}
d \omega \frac{1}{\omega^2+m^2}  A(\omega,\omega_i)
+
 2 \epsilon \int_{\Lambda}^{\infty}
d \omega \frac{1}{\omega^2+m^2}  A(\omega,\omega_i)~.
\label{eq:split}
\end{align}
The first term %in the r.h.s. of this 
of \eqref{eq:split}
is evaluated as\footnote{
The integration of the high momentum modes
is not divergent, but suppressed in one dimensional system.
} 
\begin{align}
 2 \epsilon \int_{0}^{\Lambda}
d \omega \frac{1}{\omega^2+m^2}  A(\omega,\omega_i)
\approx
 2 \epsilon 
\left( 
\int_{0}^{\infty}
d \omega \frac{1}{\omega^2+m^2}  
\right)
A(\omega=0,\omega_i) 
=
\frac{\pi \epsilon}{m} A(\omega=0,\omega_i),
\end{align}
where we neglect ${\cal O} (\frac{m}{\Lambda})$ 
and ${\cal O} (\frac{m}{|\omega_i|})$ 
suppressed terms,
and the second term is also negligible because of an
${\cal O}(\frac{m}{\Lambda})$ factor.\footnote{
Here, we have used that $A(\omega,\omega_i)$ 
and $\frac{\partial }{\partial \omega} A(\omega,\omega_i)|_{\omega=0}$
do not diverge. These follows from the fact that our model is one-dimensional system.
}

%Therefore,
This result implies that,
by identifying 
\begin{eqnarray}
 \epsilon=\frac{3!}{\pi}  \frac{m}{N^3}\cJ^2,
\end{eqnarray}
the two, four and six point correlators of our model reproduce those of the SYK model
in the large $N$ limit. %are reproduced.
In particular, for 
\begin{eqnarray}
 m \ll |\omega_i| \ll \cJ,
\end{eqnarray}
the two and four point functions are well-described by the effective action with the Schwarzian derivative proposed in \cite{Maldacena:2016hyu}.
%and  
%the saturation of the chaos bound discussed in \cite{Maldacena:2015waa} can be seen %, but,
%for $t \ll 1/m$.

\subsection{Eight-point correlation function}

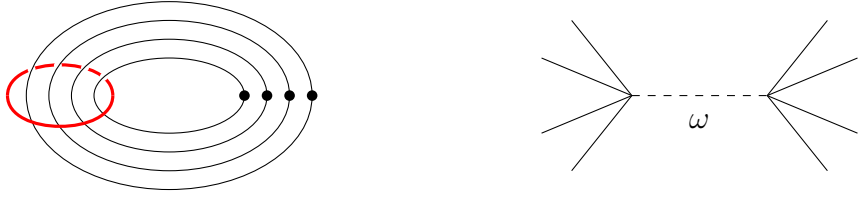
\begin{figure}
\begin{center}
\vskip .6cm
\begin{tikzpicture}
\path (-1.3,0) edge[bend left=90, color=red, very thick] (.1,0);
\fill[color=white] (-0.98,0.35) circle (1.5pt);
\fill[color=white] (-0.63,0.4) circle (1.5pt);
\fill[color=white] (-0.27,0.37) circle (1.5pt);
\fill[color=white] (-0.02,0.25) circle (1.5pt);
\draw (0.85,0) ellipse (1cm and 0.5cm);
\draw (0.85,0) ellipse (1.3cm and 0.75cm);
\draw (0.85,0) ellipse (1.6cm and 1cm);
\draw (0.85,0) ellipse (1.9cm and 1.25cm);
\fill (1.85,0) circle (2pt);
\fill (2.15,0) circle (2pt);
\fill (2.45,0) circle (2pt);
\fill (2.75,0) circle (2pt);
\path (-1.3,0) edge[bend right=90, color=red, very thick] (0.1,0);
\draw (7,0) -- (6.2,1);
\draw (7,0) -- (5.8,.5);
\draw (7,0) -- (5.8,-.5);
\draw (7,0) -- (6.2,-1);
\draw[dashed] (7,0) -- (8.8,0) node[midway, below=1mm]{$\omega$};
\draw (8.8,0) -- (9.6,1);
\draw (8.8,0) -- (10,.5);
\draw (8.8,0) -- (10,-.5);
\draw (8.8,0) -- (9.6,-1);
\end{tikzpicture}
\caption{A diagram for which the momentum $\omega$ of the $\phi$ propagator is fixed by external momenta of fermions. The left picture is in terms of dots and a bundle while the right one is an expression for the same diagram in terms of external lines and vertices.}
\label{fig:fixed}
\end{center}
\end{figure}
As we have seen, for the eight and more-than-eight point correlation functions, 
there are connected diagrams for which $\omega$ is fixed by external momenta.
% connected diagram of the leading order in $N$ contains, f
For example, let us consider the diagram depicted in (fig.~\ref{fig:fixed}).
For this diagram, the contribution of the $\phi$ %line
propagator is
\begin{eqnarray}
\int d\omega \frac{\epsilon}{\omega^2+m^2} 
\delta (\sum_j \omega_j^L+\omega)  \delta (\sum_j \omega_j^R-w) 
\sim 
 \frac{m}{(\sum_j \omega_j^L)^2+m^2} \frac{1}{N^3} \cJ^2
\delta (\sum_j \omega_j^L+\sum_j \omega_j^R) 
\label{8p} 
\end{eqnarray}
where $\sum_j \omega_j^{L(R)}$ is the sum of the momenta of 
the fermions attached to the left (right) edge of the $\phi$ line.
%However,
On the other hand, in the SYK model it is 
\begin{eqnarray}
3! \frac{\cJ^2}{N^3} 
\delta (\sum_j \omega_j^L)  \delta (\sum_j \omega_j^R).
\end{eqnarray}

Thus, the connected eight point function in the large $N$ limit
in our model is different from %the one
that %of 
in the SYK model.
However, the difference exists only 
for $(\sum_j \omega_j^L)^2={\cal O} (m)$,
which is realized by %the fine 
tuning %of
 the external momenta.
Note that a Euclidean time correlator can be obtained by the 
Fourier transform of the momentum correlator,
which includes the integration over $\sum_j \omega_j^L$.
%Thus,
Therefore, in a generic Euclidean time correlator,
(\ref{8p}) is approximated by
\begin{eqnarray}
\int d\omega \frac{\epsilon}{\omega^2+m^2} 
\delta (\sum_j \omega_j^L)  \delta (\sum_j \omega_j^R),
\end{eqnarray}
which coincides with the one in the SYK model.
In this sense, the correlation functions of our model
is almost the same as %the one
those of the SYK model.

We can use the Feynman rules in position space, instead of momentum
space.
Then, the propagator for $\phi$ in the $m \rightarrow 0$ limit 
becomes a constant:
\begin{eqnarray}
 \frac{\epsilon}{m} e^{-m| \tau|} \rightarrow 
 \frac{3!}{\pi}  \frac{1}{N^3}\cJ^2.
\label{pc}
\end{eqnarray}
With this propagator, we can easily see that 
the path-integral of the $\phi$ is equivalent to 
the disorder average.
However, 
this propagator (\ref{pc}) is not valid 
if the integration over $\tau$ for $|\tau| ={\cal O}(1/m)$
does not vanish in the limit.
As we have seen by using the propagator in momentum space, 
we expect that the naive limit (\ref{pc}) is valid
for the correlation function of the form in
(\ref{cf1}).

So far, we have considered the zero temperature result.
As we will see below, at %a
 finite temperature, 
our model is different from the SYK model with the constant random coupling $\cJ$.
On the other hand,
one could think that only fermions are excited by finite temperature while bosons are at zero temperature, 
% there exists a state in our model where the bosons are at zero temperature and the fermions are at a finite temperature,
because the bosons and fermions are weakly coupled in the large $N$ limit. Moreover, the number of the bosons is of ${\cal O} (N^4)$ and much larger than %the number 
that of the fermions, and therefore one could neglect the finite temperature effects on the bosons
%which is very weakly coupled to the fermions, 
for a sufficiently %large
long time.
If this is the case, the state corresponding to such a situation in our model will be the same as the thermal state at the temperature in the SYK model with $\cJ$.
%the coupling constant
This possibility would be interesting although we will not study it 
%this 
further in this paper. % further.

\subsection{Finite temperature}

Let us consider %the
our model at a
 finite temperature $1/\beta$. Here we take the small $m$ limit, and therefore $\beta m \ll 1$ and $m \ll |\omega_i|$.
%We %also 
%assume $\beta m \ll 1$ which corresponds to $m \ll |\omega_i| $ 
%for the external momenta.
%Then,
Since the Euclidean time direction is compactified, the momentum integration in (\ref{eq:integral}) %will be 
is now replaced 
to a discrete sum:
\begin{align}
\epsilon \frac{2 \pi}{\beta} \sum_{n} \frac{1}{(\omega_n)^2+m^2} \, A(\omega_n,\omega_i),
\label{ft1}
\end{align}
where $\omega_n=2 \pi n/ \beta$.
The dominant contribution in the small mass limit is
\begin{align}
\epsilon \frac{2 \pi}{\beta} \sum_{n} \frac{1}{(\omega_n)^2+m^2} \, A(\omega_n,\omega_i)
 \approx \frac{3!}{N^3} (\cJ_{eff})^2 \, A(0,\omega_i),
\end{align}
where $\cJ_{eff}$ is a function of $\beta m $. 
Here we note that the fermions have half-integer momenta at finite temperature,
hence there will be no singular behavior in $A(\omega, \omega_i)$ near $\omega=0$
even when there is no external lines.
For $ \beta m \ll 1$,
$J_{eff}$
is easily determined as  
%\begin{align}
%\cJ_{eff}  \rightarrow  \cJ
%\end{align}•
%for $ \beta m \rightarrow \infty$, and
\begin{align}
\cJ_{eff} =
\sqrt{\frac{2}{\beta m}} \cJ~,
\end{align}•
%for $ \beta m \rightarrow 0$ 
where only the zero mode, i.e. $n=0$,
contributes in (\ref{ft1}).

Thus, our model at the temperature $1/\beta$ reproduces the SYK model at 
the same temperature %, however,
with the coupling constant $\cJ$ 
%should be 
replaced by $\cJ_{eff}$.
%which
Note that, unlike $\cJ$, the effective coupling $\cJ_{eff}$ depends on the temperature.
This difference between our model and the SYK model
%will be 
is originated from the almost massless bosons which are 
highly excited at %the
 finite temperature.
The conformal limit, $\beta \cJ \gg 1$, in the SYK model
corresponds to $\beta \cJ_{eff} \gg 1$ 
in our model at the finite temperature.
On the other hand, as we have ignored the ${\cal O} (m)$ suppressed terms,
we need another constraint 
\begin{align}
(1 \gg) \frac{1}{\beta \cJ_{eff}} \gg m \beta,
\label{cond1}
\end{align}
where $\beta$ can be replaced by any other typical finite  scale
in the correlator.
These two inequalities can be satisfied simultaneously by choosing $\cJ$ appropriately.
Therefore, 
%This result implies that,
four point correlators of our model at %the
 finite temperature 
reproduce those of the SYK model with the temperature-dependent coupling $\cJ_{eff}$,
in the large $N$ and small $m$ limit.
In particular, 
with (\ref{cond1}),
the two and four point functions are well-described by the effective action with the Schwarzian derivative proposed in \cite{Maldacena:2016hyu}.

%Furthermore, 
Let us comment on the chaos bound proposed in \cite{Maldacena:2015waa}, which is known to be saturated by the SYK model \cite{Kitaev:2015, Polchinski:2016xgd, Maldacena:2016hyu}. This saturation of the chaos bound can be seen by studying %the
 analytic continuations of the Euclidean correlators at %the
 finite temperature.
% were used.
The Euclidean correlators at %the
 finite temperature 
in our model is the same, in the small mass limit, as the ones in the SYK model
with $\cJ$ replaced by $\cJ_{eff}$.
Thus %the
their analytic continuations 
%of the correlators 
are also the same 
for these two models.
This means that,
at the finite temperature (such that $\beta m \ll 1$),
the saturation of the chaos bound
can be seen in our model %, but,
for a sufficiently long time.

\subsection{Relations to other models }

Finally, let us comment on relations %between
to other models without disorder \cite{Michel:2016kwn}.\footnote{
Note that we use %the
a notation for %the
 fields and parameters 
which is different from the one used in \cite{Michel:2016kwn}.}
%and our model.
First, schematically, 
the Lagrangian in our model is
\begin{align}
\frac{1}{2 \epsilon} ((\dot{\phi})^2+m^2 \phi^2)+\phi \psi^4,
\end{align}
which is equivalent to 
\begin{align}
\frac{1}{2 \epsilon} (-C^2+2 \dot{\phi} C+m^2 \phi^2)+\phi \psi^4,
\label{lc}
\end{align}
where $C$ is an auxiliary field. %s.
Now we integrate out $\phi$ first, then we obtain the following Lagrangian:
\begin{align}
\frac{1}{2 \epsilon} \left(-C^2-\frac{1}{m^2} \dot{C}^2\right)-\frac{1}{m^2} \dot{C} \psi^4~,
\end{align}
where we have neglected $(\psi)^4 (\psi)^4$ term
because of the fermion anti-symmetry.
%and we introduced
In terms of $\tilde{\phi} = i C/m$, this Lagrangian is written as
\begin{align}
\frac{1}{2 \epsilon} ( (\dot{\tilde{\phi}})^2+m^2 \tilde{\phi}^2)-i \frac{1}{m} \tilde{\dot{\phi}} \psi^4,
\label{model2}
\end{align}
where the factor $i$ is necessary for the corresponding Lorentzian theory to be unitary.
 Note that we have ignored the zero modes of 
$\phi$ and $C$ in the Lagrangians
although they are important at %a
 finite temperature.
Thus, the derivative coupling model (\ref{model2}) is equivalent 
to our model except the %non-existence
absence of the zero mode.
Indeed, the propagator of $\dot{\phi}$ is
\begin{align}
-\frac{1}{m^2} \langle \dot{\phi}(\tau) \dot{\phi}(0) \rangle =
\frac{1}{m^2}\frac{\partial^2}{\partial^2 \tau} \langle \phi(\tau) \phi(0) \rangle =
\frac{\epsilon}{m} e^{- m |\tau|} \left(1-2 \frac{1}{m} \delta (\tau) \right),
\label{phid}
\end{align}
which is the same as the propagator of $\phi$ except
an extra delta-function term.
When computing correlators of $\psi$, this extra term vanishes because it %gives
 is accompanied with  $(\psi^4(\tau)) (\psi^4(\tau))$ %term
which is vanishing %by
because of the anti-commutativity of the fermions.
At %a
 finite temperature, the model %with
described by the Lagrangian (\ref{model2}) is not
%reminiscent
equivalent to the SYK model because there is no zero mode
contributions in the propagator of $\dot{\phi}$.

In \cite{Michel:2016kwn}, two quantum mechanical models without quenched disorders are briefly discussed.\footnote{See the last paragraph of Sec.~2.3 of \cite{Michel:2016kwn}.} In particular, the second model is obtained by replacing the random coupling $\cJ$ of the SYK model with the momenta of harmonic oscillators. This model is essentially equivalent to
the model described by the above Lagrangian (\ref{model2}).
%is naively a Lagrangian of the model proposed in \cite{Michel:2016kwn}.
%where the random couplings $J$ of the SYK model are replaced by the momenta of
%harmonic oscillators.
Here we note that the path-integral formulation
of a theory
is not straightforwardly obtained when the Hamiltonian has interaction terms including %the
canonical momenta.\footnote{
We also note that a correlator in the path-integral formalism
corresponds to a v.e.v. of a time ordered product operators
where derivatives are taken outsides the correlator
(sometimes this is called ${\bf T}^*$-product).
For example, for the Harmonic oscillator with $H=(\hat{x}^2+\hat{p}^2)/2$
where $\hat{p}(t) =\dot{\hat{x}} (t)$,
we see that 
\begin{align}
\langle \dot{x}(t) \dot{x}(0) \rangle 
=-\frac{\partial^2}{\partial^2 t} 
\langle 0| {\bf T} (\hat{x}(t) \hat{x}(0)) |0 \rangle 
\neq 
\langle 0|  {\bf T} (\hat{p}(t) \hat{p}(0)) |0 \rangle.
\end{align}
}
On the other hand, in the operator formalism,
an interchange between $\hat{x}$ and $\hat{p}$ 
does not change the theory except $[\hat{x},\hat{p}]=i \rightarrow [\hat{x},\hat{p}]=-i$.
Therefore, the second model in \cite{Michel:2016kwn} is essentially equivalent
to our model.

Let us also consider the first model discussed in Sec.~2.3 of \cite{Michel:2016kwn}.
%Next, the first model in \cite{Michel:2016kwn} will be described 
The model is described by the following action:
\begin{align}
\int_0^{\beta}  d\tau \left(\dot{\phi} C+ \frac{\alpha}{\beta} \phi^2+\phi \psi^4\right),
\label{ac}
\end{align}
where we have introduced $\beta$ as an IR regulator or an inverse temperature.
Because only the zero mode of $\phi$ remains after integrating out $C$,
this model is equivalent to the SYK model.
However, the Lagrangian explicitly depends on $\beta$, 
and moreover $\alpha/\beta$ vanishes in the zero temperature limit
%where 
since $\alpha = N^3/12\cJ^2$ should be fixed. Thus, this action is not suitable %as
for an action of a quantum theory.

On the other hand, the Lagrangian (\ref{lc}) of our model, at a finite temperature %with
such that $m \beta \ll 1$,
will be 
\begin{align}
\frac{1}{2 \epsilon} \left(2 \dot{\phi} C+m^2 \phi^2\right)+\phi \psi^4
=\dot{\phi} \tilde{C} + m \alpha \phi^2+\phi \psi^4,
\end{align}
where $\tilde{C}\equiv C/\epsilon$.
This is of the same form as (\ref{ac}) but does not explicitly depend on $\beta$.
This reflects the fact that the corresponding SYK model has $\beta$-dependent coupling
constant $\cJ_{eff}$ (i.e.  $\alpha_{eff}=\beta m \, \alpha$).

\section{SUSY generalization}

In \cite{Fu:2016vas}, a supersymmetric generalization of the SYK model %s are
is discussed.
In the superspace representation, the Lagrangian of the ${\cal N} =1$ model is
\begin{align}
L_{SUSY}=\int d \theta \left( \frac12 \Psi^i D_\theta \Psi^i
+i C_{ijk} \Psi^i \Psi^j \Psi^k\right), 
\label{eq:SUSY}
\end{align}
where
$\Psi^i=\psi(\tau)+\theta b(\tau)$,
$D_\theta=\partial_\theta+\theta \partial_\tau$ 
and $C_{ijk}$ is a Gaussian random coupling with 
\begin{align}
\bar{C_{ijk}^2}=\frac{2J}{N^2}, \,\,\, \bar{C_{ijk}}=0.
\end{align}

We now promote $C_{ijk}$ to %the
a dynamical field
as follows.
First, we introduce the following bosonic and fermionic superfields:
$X_{ijk}=C_{ijk}(\tau)+\theta \eta_{ijk}(\tau)$ 
and $\Xi_{ijk}=\xi_{ijk}(\tau) +\theta F_{ijk}(\tau)$.
Then, the Lagrangian of our model is $L_{SUSY}+L_X$ where 
\begin{align}
L_X=\int d \theta  \frac{1}{2 \epsilon} \left(D^2_\theta X_{ijk} D_\theta X_{ijk}
+\Xi_{ijk} D_\theta \Xi_{ijk} 
+m \Xi_{ijk} X_{ijk}\right).
\end{align}
Note that we need to introduce the fermionic partners of $C$.
In terms of the superfield $\Psi^i$, the action \eqref{eq:SUSY} is of the same form as the action of the $q=3$ SYK model,\footnote{Here $q$ stands for the number fermions involved in the interaction term, as in \cite{Maldacena:2016hyu}.} a similar discussion on the large $N$ limit as in the previous section may be available. Therefore we expect that diagrams including ``bubbles'' are sub-leading in the large $N$ limit, as in the bosonic case.
 %In this superspace representation, the action has the form 
%with $q=3$ of the non-SUSY model discussed in the previous sections.
%Thus, the $N$-dependence in the large $N$ limit 
%may be applicable for this supersymmetric model
%and we expect that diagrams includes "bubbles" 
%are not leading order in $N$. 

Now we consider the massless limit of the propagator at zero temperature 
of this SUSY model. We take $\epsilon \sim m J/N^2$ so that the propagators of the bosonic components $C$ become constants as we have seen before in (\ref{pc}).
%For the bosonic component, $C$, it becomes the constant 
%as we have seen before in (\ref{pc}).
%Thus, we will take $\epsilon \sim m J/N^2$.
On the other hand, for the fermions $\eta$,
the propagator in position space becomes
\begin{align}
\epsilon \frac12 {\rm sgn} (\tau)~,
\end{align}
which is of $\cO(m)$ and therefore suppressed compared with the bosonic propagator in the massless limit.
%This is ${\cal O}(m)$ suppressed compared with the bosonic one,
%and therefore can be ignored in the small $m$ limit.
%There seems to be a conflict between the SUSY and this difference, however, 
This difference between bosons and fermions does not contradict with supersymmetry because the fermion propagator is related by supersymmetry to the derivative of 
the bosonic one, i.e., %$\frac{\epsilon}{m} e^{-m |\tau|}$ as 
\begin{align}
\frac{\partial}{\partial \tau}\frac{\epsilon}{m} e^{-m |\tau|} =    \frac{\partial}{\partial \tau} \left( 
\frac{\epsilon}{m} - \epsilon |\tau|+ \cdots 
\right)~,
\end{align}
where only the sub-leading part in the massless limit remains.
Therefore, in the massless limit, the fermionic partner of $C$ decouples from the $\Psi$. Moreover, since $X$ and $\Xi$ are decoupled from each other in the massless limit,
 the path-integrations over the dynamical fields $X,\Xi$ reduce to %becomes 
Gaussian random couplings as in the non-SUSY model discussed in the previous sections.
Therefore, our SUSY model described by $L_{SUSY}+L_{X}$ 
is essentially equivalent to the original SUSY generalization of 
the SYK model in the large $N$ and massless limit.

\newpage

\section*{Acknowledgements}
The authors thank Masamichi Miyaji, Kazuma Shimizu, Tadashi Takayanagi and Kento Watanabe
for illuminating discussions.
The work of T.~N. is partially supported by the Yukawa Memorial Foundation (T.~N. was the Yukawa Rsearch Fellow), and by JSPS Grant-in-Aid for Scientific Research (B) No. 16H03979.

\bibliographystyle{tref}
\bibliography{ref}

\end{document}